\documentclass[conference,10pt]{IEEEtran}
\usepackage{epsfig,rotating,setspace,latexsym,amsmath,epsf,amssymb,amsfonts,bm,theorem, epstopdf,cite,authblk, bbm, caption, subcaption, multirow, enumitem}
\usepackage[ruled,vlined]{algorithm2e}

\usepackage{color}

\IEEEoverridecommandlockouts
\allowdisplaybreaks

\begin{document}
\title{Adversarial Attack and Defense for LoRa Device Identification and Authentication via Deep Learning}	
	
\author{Yalin E. Sagduyu and Tugba Erpek  \thanks{Yalin Sagduyu and Tugba Erpek are with Nexcepta, Gaithersburg, MD, USA.  Email: ysagduyu@nexcepta.com, terpek@nexcepta.com.} \thanks{A preliminary version of this paper was partially presented at IEEE Military Communications Conference (MILCOM), 2023 \cite{sagduyuLoRAMILCOM}.}}
	
\maketitle
\begin{abstract}
LoRa provides long-range, energy-efficient communications in Internet of Things (IoT) applications that rely on Low-Power Wide-Area Network (LPWAN) capabilities. Despite these merits, concerns persist regarding the security of LoRa networks, especially in situations where device identification and authentication are imperative to secure the reliable access to the LoRa networks. This paper explores a deep learning (DL) approach to tackle these concerns, focusing on two critical tasks, namely (i) identifying LoRa devices and (ii) classifying them to legitimate and rogue devices. Deep neural networks (DNNs), encompassing both convolutional and feedforward neural networks, are trained for these tasks using actual LoRa signal data. In this setting, the adversaries may spoof rogue LoRa signals through the kernel density estimation (KDE) method based on legitimate device signals that are received by the adversaries. Two cases are considered, (i) training two separate classifiers, one for each of the two tasks, and (ii) training a multi-task classifier for both tasks. The vulnerabilities of the resulting DNNs to manipulations in input samples are studied in form of untargeted and targeted adversarial attacks using the Fast Gradient Sign Method (FGSM). Individual and common perturbations are considered against single-task and multi-task classifiers for the LoRa signal analysis. To provide resilience against such attacks, a defense approach is presented by increasing the robustness of classifiers with adversarial training. Results quantify how vulnerable LoRa signal classification tasks are to adversarial attacks and emphasize the need to fortify IoT applications against these subtle yet effective threats.

\end{abstract}

\section{Introduction}
A cost-effective communication solution is provided by \emph{LoRa} for extended connectivity range, reduced power usage, prolonged battery life, and scalability. This makes LoRa well-suited for diverse applications of Internet of Things (IoT) that rely on Low Power Wide Area Networks (LPWANs), including asset tracking, industrial automation, surveillance, smart city infrastructure, smart home systems, and supply chain management \cite{bor2016lora, sinha2017survey, zourmand2019internet, LoRatactical1,10122600}. Although LoRa presents various benefits, it is also subject to \emph{security} threats. A significant concern involves unauthorized access, as adversaries may attempt to breach the LoRa network, intercept or tamper with data, and prevent reliable device communications. LoRa security vulnerabilities are studied in \cite{10.1145/3543856}. LoRa signals are susceptible to interception anoverd eavesdropping over the air. In potential replay attacks, adversaries may capture signals from legitimate LoRa devices and resend them to deceive the LoRa network \cite{7985777,9619847}. Adversaries may also engage in spoofing or impersonation of legitimate LoRa device signals with the intent of gaining unauthorized access, injecting malicious data, or undermining network operations. Consequently, it is crucial to assess the attack surface associated with LoRa and design defense mechanisms \cite{9072101}.

\emph{Signal classification} is essential in identifying LoRa devices and detecting rogue device transmissions. Enabled by recent computational and algorithmic advancements, \emph{deep learning} (DL) is known to effectively extract advanced features from raw signals, facilitating precise and efficient classification of wireless signals \cite{erpek2020deep, west2017deep, shi2019deep}. To that end, deep neural networks (DNNs) are promising to reliably perform LoRa signal analysis tasks. Through training on real-world I/Q data collected from LoRa devices, the DNNs can capture subtle signal characteristics and enhance classification accuracy \cite{Lora1, robyns2017physical, elmaghbub2021lora, al2021deeplora, shen2022towards, 9621015,9093371}. The state of the art in the area of DL-based radio frequency fingerprinting identification for LoRa devices is surveyed in \cite{s24134411}. 

There are potentially \emph{multiple tasks} to perform when analyzing wireless signals \cite{jagannath2021multi}. In this paper, we focus on \emph{two signal classification tasks} within LoRa networks: (i) Task 1: distinguishing among LoRa devices, (ii) Task 2: detecting spoofed signals that imitate LoRa signal characteristics. We adopt convolutional neural network (CNN) and feedforward neural network (FNN) architectures to implement classifiers for these tasks. For signal spoofing, adversaries generate \emph{synthetic signals} to deceive the spectrum monitors and potentially bypass the physical-layer authentication mechanisms \cite{shi2019generative, shi2020generative}. In this paper, we apply the \emph{Kernel Density Estimation} (KDE) method to generate synthetic signals by learning probability distributions of legitimate LoRa signals. We show that the DNNs achieve high accuracy for the two classification tasks that we consider for LoRa networks. For that purpose, we consider either two single-task classifiers, one for each task, or a \emph{multi-task classifier} that trains a common model for both tasks, leveraging shared information across tasks.

The intricate decision space inherent in classification of wireless signals exhibits sensitivity to variations in input samples during test (inference) time, rendering it susceptible to \emph{adversarial (evasion) attacks} \cite{sadeghi2018adversarial, kim2021channel, adesina2022adversarial, sagduyu2020wireless}. Adversaries can determine subtle yet harmful perturbations in the input signals during test time with the aim of misleading the DNN models used for LoRa device identification and rogue signal detection, compromising the advanced authentication mechanisms that rely on RF fingerprinting and facilitating unauthorized access to the LoRa network.

Adversarial attacks against a DL task of classifying LoRa signals have been investigated in \cite{sagduyuLoRAMILCOM, 10278927} using different DNN types and real LoRa signal datasets. To the best of our knowledge, adversarial attacks against multi-task classifiers for LoRa signal analysis have not been investigated earlier. In this paper, we consider two classification tasks (LoRa device identification and rogue LoRa signal detection) and investigate their individual and joint DNN models as single and multi-task classification problems. First, we explore adversarial attacks for these models against both single-task and multi-task classifiers and evaluate their impact on different tasks. Then, we develop defense mechanisms based on \emph{adversarial training}. Our study encompasses \emph{untargeted attacks}, which seek to fool classifiers into making error without specifying a target label, and \emph{targeted attacks}, which intend to compel the model to misclassify input samples to a target label. Fast gradient sign method (FGSM) is adopted to generate adversarial perturbations. This method involves computing the gradient of the loss function concerning the input data and subsequently adding perturbation to the input samples in the direction of the gradient sign with the intent of manipulating the model's loss \cite{goodfellow2015explaining}.

Earlier work mainly focuses on IoT device classification using legitimate LoRa signals \cite{elmaghbub2021lora}. We provide a comprehensive analysis that includes design of LoRa device type and rogue and legitimate device classification models and design of (both untargeted and targeted) adversarial attacks on single and multi-task signal classification and corresponding defense mechanisms. The performance is evaluated using over-the-air captured LoRa signal with realistic channel conditions.

\begin{figure}[t]
   
	\centering
	\includegraphics[width=0.935\columnwidth]{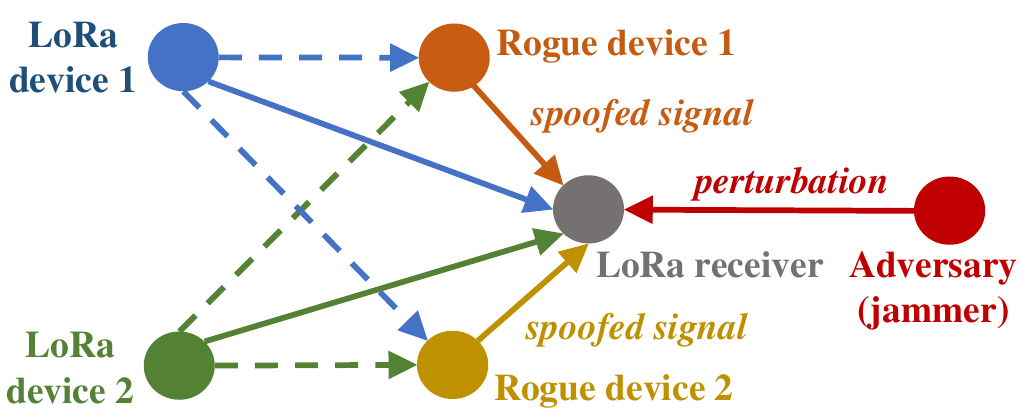}
	\caption{{ \footnotesize System model.}}
	\label{fig:LoRa_system}
 \vspace{-0.2cm}
\end{figure}

As shown in Fig.~\ref{fig:LoRa_system}, we consider two LoRa devices engaged in transmissions and two rogue devices that mimic the transmissions of legitimate devices. The LoRa receiver has two tasks, namely classifying its received signals to Device 1 vs. Device 2  (Task 1) and classifying its received signals to legitimate vs. rogue signals (Task 2).

Concurrently, an adversary broadcasts a common perturbation signal as an adversarial attack to fool classifiers for both tasks. We evaluate the impact of this adversarial attack and explore \emph{transferability} \cite{papernot2016transferability} in adversarial attacks by assessing whether adversarial examples that are intended to deceive the model for one task can also mislead models for the other task. 
We show that there may be a significant decline in attack performance when the model undergoing the attack does not match the model for which the perturbation was initially determined. To address this challenge, a single shared perturbation is generated by combining the gradients of the loss function across different classifiers. This hybrid attack remains effective irrespective of the classifier model and the type of DNN being attacked. 

We extend the analysis to adversarial attacks on a \emph{multi-task classifier} that trains a common model for the two tasks. By crafting a single (common) perturbation, we show that the multi-task classifier becomes susceptible to adversarial attacks such that the accuracy of both tasks can significantly drop.

Subsequently, we study the \emph{defense} to protect the LoRa network against adversarial attacks. \emph{Adversarial training} is used to augment the training data by incorporating adversarial inputs in addition to the original samples, thereby fortifying the trained models against adversarial attacks \cite{goodfellow2015explaining}. We show that this defense renders adversarial attacks ineffective with only a small effect on the classification accuracy for clean data samples that have no perturbation added.

The novelty of technical contributions in this paper can be summarized as follows:
\begin{enumerate}
    \item Earlier work mainly focuses on IoT device classification using legitimate LoRa signals, studied such as in \cite{elmaghbub2021lora}. In this paper, we present a new approach to generate rogue LoRa signals and developed classifiers for multiple tasks, namely for both IoT device classification and rogue vs. legitimate device classification.
    \item We investigate the performance of both individual and joint DNN models as single and multi-task classification to classify LoRa device type and legitimate and rogue devices as opposed to only single task classification in the earlier work. We combined joint training of different tasks through the optimization approach of multi-task learning.
    \item Prior work has studied adversarial attacks against attacks such as modulation classification and spectrum sensing \cite{adesina2022adversarial}. In this paper, we develop adversarial attacks (both untargeted and targeted) against different LoRa signal analysis tasks. 
    \item Prior work on adversarial attacks against wireless systems has focused on single-task problems. In this paper, we develop adversarial attacks against single and multi-task models for LoRA signal analysis. To that end, we studied the transferrability of attacks from one task to another and developed attacks against multi-task learning. Finally, while some prior work has been limited to attack studies in wireless settings, we also present a defense mechanism based on adversarial training against adversarial attacks on single-task and multi-task LoRa signal analysis.   
\end{enumerate}

The rest of this paper is organized as follows. 
Sec.~\ref{sec:signal} examines LoRa signal analysis using single-task classifiers. Sec.~\ref{sec:multitask} presents the multi-task classification approach. Sec.~\ref{sec:attack} studies adversarial attacks that are launched against different tasks of LoRa signal classification. Sec.~\ref{sec:defense} explores the defense mechanism. Sec.~\ref{sec:conclusion} presents concluding remarks.

\section{LoRa Signal Classification} \label{sec:signal}
 We utilize the I/Q data from \cite{elmaghbub2021lora}, sourced from authentic LoRa (Pycom IoT) devices acting as the transmitters and software-defined radios (SDRs), specifically, USRP B210 radios, acting as the receivers. In an outdoor setting, signals of LoRa devices operating at a center frequency of 915MHz have been captured with sampling rate of 1MS/s. The LoRa configuration comprises Raw-LoRa mode with 125kHz as the channel bandwidth, 7 as the spreading factor, 8-preamble, 20 dBm as the transmit power, 4/5 as the coding rate, and 5 meter as the distance between the transmitter and the receiver. After reading the I/Q samples from the files, we have directly used them as input to the DNN classifiers. We have not used any data preprocessing on the data. We also have not used any data augmentation techniques since there are sufficient number of samples in the dataset to achieve high classification accuracy. The dataset is balanced across classes, and we have not observed any biases in the data. 
We process each data sample to become an input of dimension (2,32) to the DNN, corresponding to 32 I and Q pairs. This way, we generate a dataset of 5000 samples, with 80\% and 20\% allocated for training and testing, respectively. 

In this paper, we consider two tasks of classifying LoRa signals, Task 1: classifying the received signals to  `Device 1' and `Device 2' depending on whether the transmissions are originated from LoRa Device 1 or Device 2, and Task 2: classifying the received signals to `legitimate' vs. `rogue' depending on whether the transmissions are originated from legitimate or rogue LoRa devices. Classifiers for Task 1 and Task 2 are referred to as Classifier 1 and Classifier 2.

\subsection{Spoofed Signal Generation by Rogue LoRa Devices} \label{sec:rogue}
The adversary employs KDE to generate signals for LoRa device spoofing. KDE calculates the probability density function (PDF) of legitimate LoRa signals received by rogue devices, and can be used to produce synthetic signals that have similar statistical properties as original signals. Transmitting these synthetic signals may help the adversary infiltrate the authentication mechanism and gain access to the LoRa network. 

\begin{figure}[t]
\vspace{-0.4cm}
\captionsetup[subfigure]{justification=centering}
\centering
\begin{subfigure}[b]{0.24\columnwidth} 
\centering
\includegraphics[width=0.935\columnwidth]{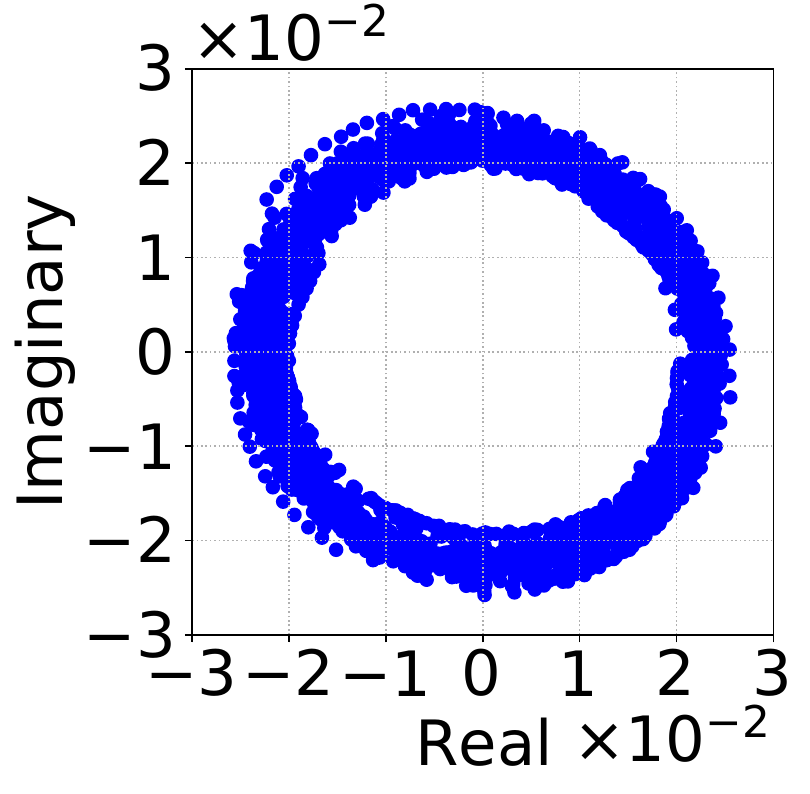}
\caption{Legitimate Device 1.}
\label{fig:Imagetrigger1}
\end{subfigure}
\begin{subfigure}[b]{0.24\columnwidth}
\centering
\includegraphics[width=0.935\columnwidth]{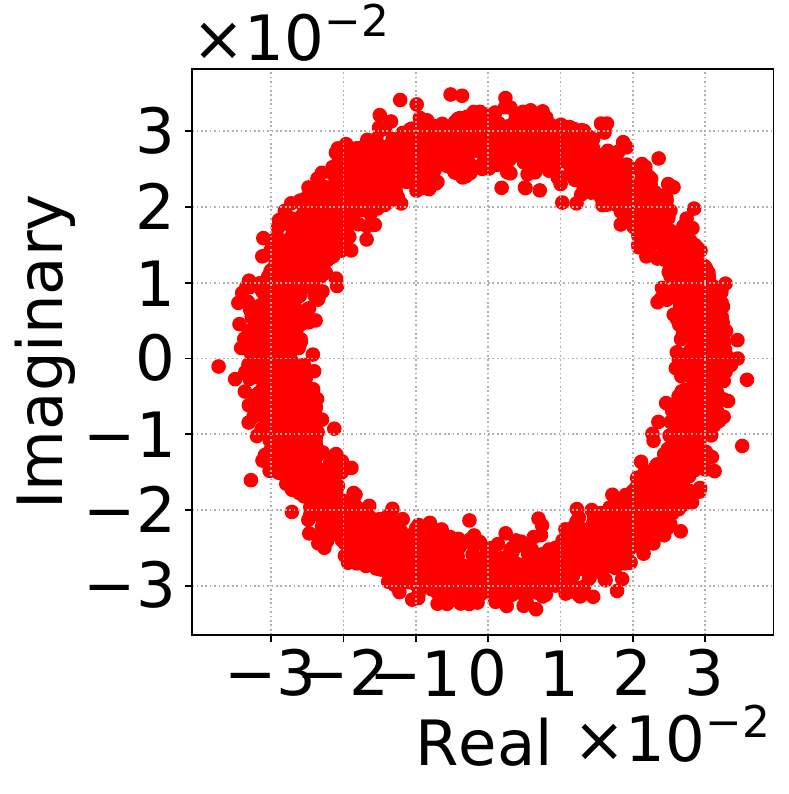}
\caption{Rogue Device 1.}
\label{fig:Imagerecons1}
\end{subfigure}
\begin{subfigure}[b]{0.24\columnwidth}
\centering
\vspace{0.3cm}
\includegraphics[width=0.935\columnwidth]{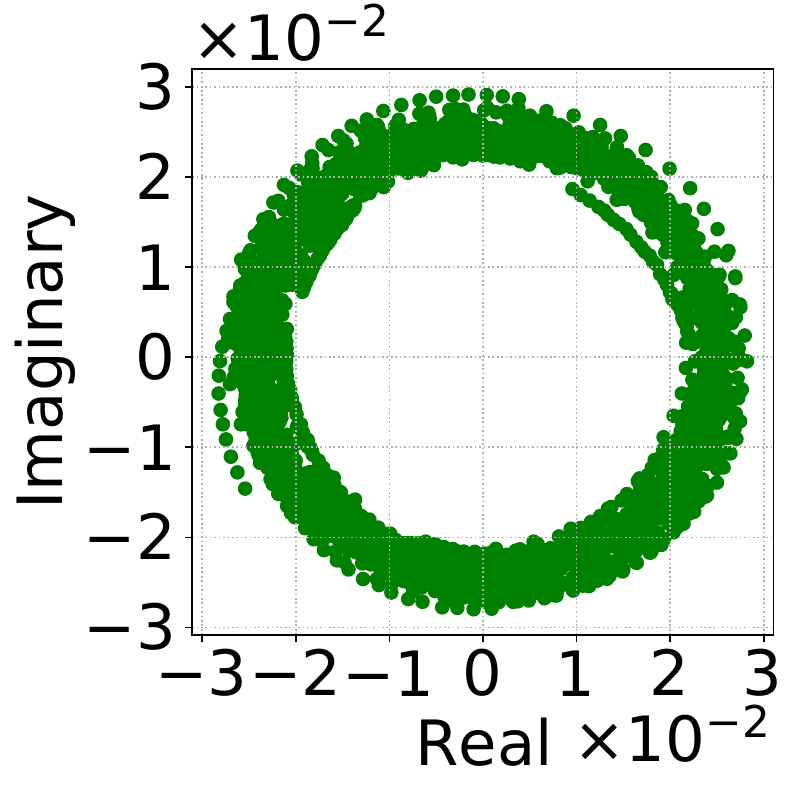}
\caption{Legitimate Device 2.}
\label{fig:Imagetrigger2}
\end{subfigure}
\begin{subfigure}[b]{0.24\columnwidth}
\centering
\vspace{0.3cm}
\includegraphics[width=0.935\columnwidth]{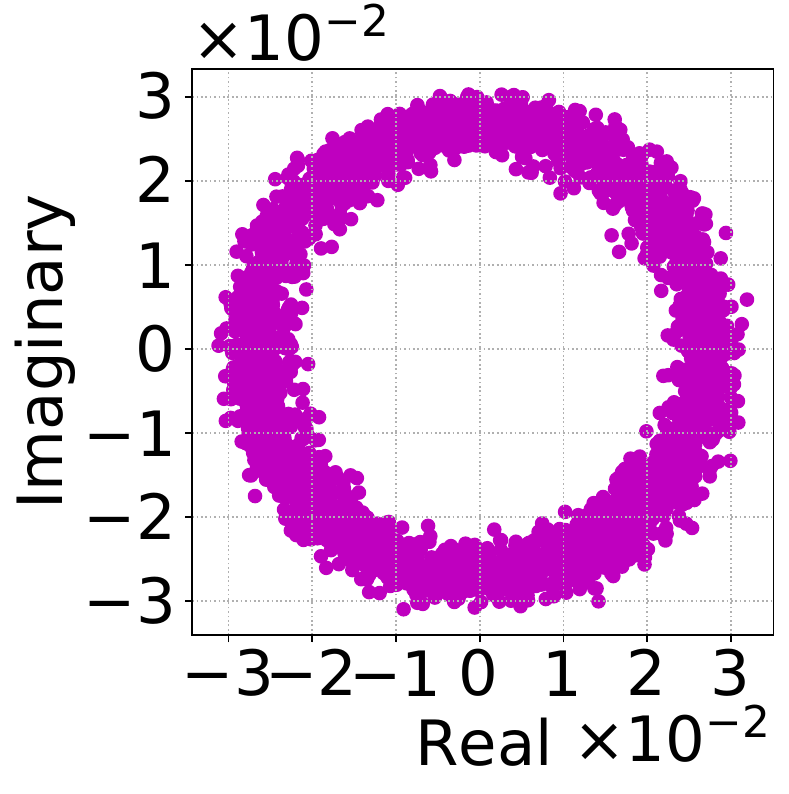}
\centering
\caption{Rogue Device 2.}
\label{fig:Imagerecons2}
\end{subfigure}
\caption{Legitimate and rogue device signal constellations for 100 samples, each consisting of 32 I/Q signals.} \label{fig:constellation}
\vspace{-0.3cm}
\end{figure}

KDE involves applying kernel smoothing techniques to estimate probability density using the observed samples. During this procedure, a kernel function is centered at each observed data point, and its effect on the PDF estimate is computed for the given kernel and bandwidth. The form of the kernel is determined by the kernel function, while the bandwidth determines its width, influencing the smoothness of the estimate and achieving a balance between capturing intricate details and preventing excessive smoothing of the PDF estimate. The involvement of the kernel entails a proportionally adjusted version of the kernel function assessed at the precise data point. Subsequently, the specific contributions from the kernels are aggregated to derive the final PDF estimation, guaranteeing the smoothness and continuity of the estimated PDF. Utilizing KDE on observed data yields an estimate that represents the inherent probability distribution. With $(x_1, x_2, \ldots, x_n)$ representing the set of observed data points, the computation of the KDE estimates $f(x)$ for the PDF at a particular point $x$ is given as
\begin{equation}
f(x) = \frac{1}{nh} \sum_{i=1}^{n} K\left(\frac{x - x_i}{h}\right), \label{eq:KDE}
\end{equation}
where $K$ denotes the kernel function, $h$ denotes the bandwidth, $x_i$ denotes individual observed data points, and $n$ denotes the total count of observed data points.
The KDE estimate $f(x)$ for a specific point $x$ is computed by combining the scaled contributions of the kernel function, each assessed at the observed data points $x_i$. Applying the scaling factor $\frac{1}{nh}$ guarantees that the estimated PDF integrates to $1$ over range of $x$. For signal spoofing, we utilize the Gaussian distribution as the kernel function with a bandwidth set to $10^{-3}$. Rogue Device 1, mimicking legitimate Device 1, transmits with a difference exceeding 2dB from the legitimate counterpart, while Rogue Device 2, emulating legitimate Device 2, transmits with a difference of less than 1dB from the authentic Device 2. Additionally, each rogue device exhibits a phase difference of up to $\frac{\pi}{30}$ degrees from the corresponding legitimate device. The constellation for both the legitimate and rogue Devices 1 and 2 is depicted in Fig.~\ref{fig:constellation}.

We assess the fidelity of KDE-generated synthetic data with the Jensen-Shannon divergence (JSD). JSD measures the similarity of two probability distributions. When $P_i$ and $\hat P_i$ denote the probability distributions for legitimate and rogue device $=1,2$, the JSD is given by 
\begin{equation}
\textit{JSD}(P_i, \hat P_i ) = \frac{1}{2}  \left(\textit{KL}(P_i || M_i ) +  \textit{KL}(\hat P_i || M_i ) \right),
\end{equation}
where $M_i = \frac{1}{2} \bigl(P_i+\hat P_i \bigl)$ and the Kullback-Leibler (KL) divergence between probability distributions $P$ and $Q$ on the same sample space $\mathcal{X}$ is given by
\begin{equation}
\textit{KL}(P||Q) = \sum_{x \in \mathcal{X}} P(x) \log \left( \frac{P(x)}{Q(x)}\right).
\end{equation}
where $P$ is the true and $Q$ is the model probability distribution. The JSD is from the range of 0 to 1, where 0 denotes identical distributions, and 1 signifies complete dissimilarity. Hence, a reduced JSD value indicates greater fidelity between the synthetic data produced by the KDE and the original data. 
Regarding the test data, the JSD values are $\textit{JSD}(P_1, \hat P_1 ) = 0.0092$ for Device 1 and $\textit{JSD}(P_2, \hat P_2 ) = 0.0099$ for Device 2, leading to an average JSD of $0.0096$ for both devices. This small  JSD indicates that the KDE-generated synthetic signals closely reflect the statistical features of the original LoRa signals (channel and hardware properties), highlighting the high accuracy in spoofing LoRa signals by rogue devices. 

\subsection{Deep Learning for LoRa Signal Classification}
For each signal classification task, we consider either a CNN or an FNN classifier. Table~\ref{tab:NNArch} shows the DNN architectures. These FNN and CNN classifiers feature 6,522 and 70,074 trainable parameters, respectively. The training process minimizes the categorical cross-entropy as the loss function by using the Adam optimizer \cite{adam}. We aim at optimally selecting the hyperparameters of the DNN classifiers to achieve a high classification accuracy, i.e., higher than 85\%, for both of the DNN architectures and both of the tasks by increasing the number of neural network layers and number of neurons in each layer in different runs. We gradually increase these hyperparameters until achieving the best performance without overfitting, while keeping the deep neural network models small for resource-constrained implementation. During training, we set the number of epochs as $50$ and during evaluation we used the model parameters that provide the highest classification accuracy.

\begin{table}[ht]
\footnotesize
    \centering
    \caption{DNN architectures.}
    \label{tab:NNArch}
    \begin{subtable}[t]{0.5\textwidth}
    \centering
    \caption{DNN type: CNN.}
    \begin{tabular}{l||l}
    \hline
    Layers & Properties \\ \hline
       Conv2D & filters = 32, kernel size = (1,3), \\ & activation function = ReLU\\
       Flatten & -- \\
       Dense & size = 32, activation function = ReLU\\
       Dropout & dropout rate = $0.1$ \\
       Dense & size = 8, activation function = ReLU \\
       Dropout & dropout rate = $0.1$\\
       Dense & size = 2, activation function = SoftMax \\ \hline
    \end{tabular}
    \vspace{0.25cm}
    \end{subtable}
    \begin{subtable}[t]{0.5\textwidth}
    \centering
    \caption{DNN type: FNN.}
    \begin{tabular}{l||l}
    \hline
    Layers & Properties \\ \hline
       Dense  & size = 64, activation function = ReLU \\
       Dropout & dropout rate = $0.1$ \\
       Dense & size = 32, activation function = ReLU \\
       Dropout & dropout rate = $0.1$ \\
       Dense & activation function = ReLU \\
       Dropout & dropout rate = $0.1$ \\
       Dense & size = 2, activation function  = SoftMax \\ \hline
    \end{tabular}
    \end{subtable} 
    
\end{table}

Tables \ref{tab:threstasksCNN} and \ref{tab:threstasksFNN} show the classification accuracy of Tasks 1 and 2, when the DNN type is CNN and FNN, respectively. Note that Task 1 is performed by using data from either the combination of legitimate and rogue device transmissions, only legitimate device transmission, or only rogue device transmissions. Predicted and true labels, denoted by $L_{\text{pred}}$ and $L_{\text{true}}$, are from \{`Device 1', `Device 2'\} for Task 1 and \{`legitimate', `rogue'\} for Task 2.  While the accuracy is consistently high, the accuracy is higher for Task 2 compared to Task 1. Specifically, for Task 1, the accuracy diminishes when incorporating transmissions from both legitimate and rogue devices (a more challenging scenario), in contrast to utilizing transmissions from only legitimate or rogue devices.

Supervised learning algorithms require labeled data for each output class during training. They do not offer the inherent capability to correctly classify a new device type that has not been seen during training. In this paper, we assume that the device types do not change during the operation time. 
Adaptation of deep learning in the classification process can be achieved through continual learning. Continual learning methods such as elastic weight consolidation (EWC) \cite{8935684} or PackNet \cite{davaslioglu2024continualdeepreinforcementlearning} can be adopted to enable the DNN classifiers to classify new device types without training from scratch.

\begin{table}[ht]
\footnotesize
    \centering
    \caption{Single-task classification accuracy with CNN.}
    \label{tab:threstasksCNN}
    
    \begin{subtable}[t]{0.4\textwidth}
    \centering
    \caption{Task 1 of classifying `Device 1' vs. `Device 2' using both legitimate and rogue device transmissions.}
    \begin{tabular}{l||l}
    \hline
    Pr\big($L_{\text{pred}}$ = $L_{\text{true}}$\big)  & 0.9060 \\ 
    Pr\big($L_{\text{pred}}$ = `Device 1' $\vert$ $L_{\text{true}}$ = `Device 1'\big) & 0.8946 \\ 
     Pr\big($L_{\text{pred}}$ = `Device 2' $\vert$ $L_{\text{true}}$ = `Device 2'\big) & 0.9175  \\ \hline 
    \end{tabular}
    \vspace{0.25cm}
    \end{subtable} 

 \begin{subtable}[t]{0.4\textwidth}
    \centering
    \caption{Task 1 of classifying `Device 1' vs. `Device 2' using only legitimate device transmissions.}
    \begin{tabular}{l||l}
     \hline
    Pr\big($L_{\text{pred}}$ = $L_{\text{true}}$\big)   & 0.9330 \\ 
   Pr\big($L_{\text{pred}}$ = `Device 1' $\vert$ $L_{\text{true}}$ = `Device 1'\big) & 0.9524\\ 
    Pr\big($L_{\text{pred}}$ = `Device 2' $\vert$ $L_{\text{true}}$ = `Device 2'\big) & 0.9133 \\ \hline 
    \end{tabular}
    \vspace{0.25cm}
    \end{subtable} 

    \begin{subtable}[t]{0.4\textwidth}
    \centering
    \caption{Task 1 of classifying `Device 1' vs. `Device 2' using only rogue device transmissions.}
    \begin{tabular}{l||l}
     \hline
    Pr\big($L_{\text{pred}}$ = $L_{\text{true}}$\big)   & 0.9300 \\ 
   Pr\big($L_{\text{pred}}$ = `Device 1' $\vert$ $L_{\text{true}}$ = `Device 1'\big) & 0.9226 \\ 
    Pr\big($L_{\text{pred}}$ = `Device 2' $\vert$ $L_{\text{true}}$ = `Device 2'\big) & 0.9375 \\ \hline 
    \end{tabular}
    \vspace{0.25cm}
    \end{subtable} 

    \begin{subtable}[t]{0.4\textwidth}
    \centering
    \caption{Task 2 of classifying `legitimate' vs. `rogue' devices using Device 1 and Device 2 transmissions.}
    \begin{tabular}{l||l}
    \hline
    Pr\big($L_{\text{pred}}$ = $L_{\text{true}}$\big)  & 0.9755 \\ 
    Pr\big($L_{\text{pred}}$ = `legitimate' $\vert$ $L_{\text{true}}$ = `legitimate'\big) & 0.9800 \\ 
    Pr\big($L_{\text{pred}}$ = `rogue' $\vert$ $L_{\text{true}}$ = `rogue'\big) & 0.9710 \\ \hline
    \end{tabular}
    \vspace{0.5cm}
    \end{subtable}

    \caption{Single-task classification accuracy with FNN.}
    \label{tab:threstasksFNN}
    
    \begin{subtable}[t]{0.4\textwidth}
    \centering
    \caption{Task 1 of classifying `Device 1' vs. `Device 2' using legitimate and rogue device transmissions.}
    \begin{tabular}{l||l}
    \hline
    Pr\big($L_{\text{pred}}$ = $L_{\text{true}}$\big)  & 0.8956 \\ 
    Pr\big($L_{\text{pred}}$ = `Device 1' $\vert$ $L_{\text{true}}$ = `Device 1'\big) & 0.8737 \\ 
    Pr\big($L_{\text{pred}}$ = `Device 2' $\vert$ $L_{\text{true}}$ = `Device 2'\big) & 0.9175 \\ \hline 
    \end{tabular}
    \vspace{0.25cm}
    \end{subtable} 

    \begin{subtable}[t]{0.4\textwidth}
    \centering
    \caption{Task 1 of classifying `Device 1' vs. `Device 2' using only legitimate device transmissions.}
    \begin{tabular}{l||l}
     \hline
    Pr\big($L_{\text{pred}}$ = $L_{\text{true}}$\big)  & 0.9180 \\ 
    Pr\big($L_{\text{pred}}$ = `Device 1' $\vert$ $L_{\text{true}}$ = `Device 1'\big) & 0.9345 \\ 
    Pr\big($L_{\text{pred}}$ = `Device 2' $\vert$ $L_{\text{true}}$ = `Device 2'\big) & 0.9012 \\ \hline 
    \end{tabular}
    \vspace{0.25cm}
    \end{subtable}

    \begin{subtable}[t]{0.4\textwidth}
    \centering
    \caption{Task 1 of classifying `Device 1' vs. `Device 2' using only rogue device transmissions.}
    \begin{tabular}{l||l}
     \hline
    Pr\big($L_{\text{pred}}$ = $L_{\text{true}}$\big)  & 0.9190 \\ 
    Pr\big($L_{\text{pred}}$ = `Device 1' $\vert$ $L_{\text{true}}$ = `Device 1'\big) & 0.9102 \\ 
    Pr\big($L_{\text{pred}}$ = `Device 2' $\vert$ $L_{\text{true}}$ = `Device 2'\big) & 0.9274 \\ \hline 
    \end{tabular}
    \vspace{0.25cm}
    \end{subtable} 

    \begin{subtable}[t]{0.4\textwidth}
    \centering
    \caption{Task 2 of classifying `legitimate' vs. `rogue' devices using Device 1 and Device 2 transmissions.}
    \begin{tabular}{l||l}
    \hline
    Pr\big($L_{\text{pred}}$ = $L_{\text{true}}$\big)  & 0.9700 \\ 
    Pr\big($L_{\text{pred}}$ = `legitimate' $\vert$ $L_{\text{true}}$ = `legitimate'\big) & 0.9771 \\
   Pr\big($L_{\text{pred}}$ = `rogue' $\vert$ $L_{\text{true}}$ = `rogue'\big) & 0.9629 \\\hline
    \end{tabular}
    \end{subtable}
    
\end{table}

\section{Multi-task Learning for LoRa Signal Classification} \label{sec:multitask}
In traditional single-task learning, a separate model is trained independently for each task, as discussed in Sec.~\ref{sec:signal}. Multi-task learning (MTL) aims to jointly train a model on multiple related tasks, leveraging shared information across tasks to improve the overall performance. 
MTL optimizes a joint objective function that encompasses the loss functions of all the tasks involved and trains a model to learn shared representations that capture essential features for both classification tasks. The benefits of MTL include shared feature learning, transfer of knowledge, and regularization effect. We apply MTL for Tasks 1 and 2 considered in Sec.~\ref{sec:signal}. When we consider each Task $i = 1,2$, separately, the objective of DL is to minimize a loss function for Task $i$,  given by
\begin{equation}
L_{i}\big(x,y_i,\theta_i\big) = \frac{1}{N_{i}} \sum_{j=1}^{N_{i}} \ell_{i}\big(y_{i,j}, f_{\theta_i}(x_{j})\big),
\end{equation}
where $\theta_i$ is the set of model parameters, $N_{i}$ is the number of samples, $y_{i,j}$ is the $j$th element of the ground truth labels, $x_{j}$ is the $j$th element of input features, $f_{\theta_i}$ is the neural network with parameters $\theta_i$, and $\ell_{i}$ is loss functions for Task $i$.
In MTL, we introduce task-specific weighting factors denoted as $w_{1}$ and $w_{2}$ and construct the loss function to be minimized as follows:
\begin{equation}
L_{\text{multi}}\big(x,y_1,y_2, \theta\big) = \sum_{i=1}^2 w_i  L_i \big(x,y_i,\theta\big),
\end{equation}
where $w_{i}$ serves as weight controlling the importance of Task $i$ in the overall learning process, $0 \leq w_i \leq 1$ for $i=1,2$ and $w_1+w_2=1$. 

The DNN architecture for MTL consists of three parts, a common branch (shared layer) followed two individual branches (task-specific layers), one for each of Tasks 1 and 2. The common branch is the first layer of the DNN architecture shown in Table \ref{tab:NNArch} and each of individual branches consists of the rest of layers of the DNN architecture shown in Table \ref{tab:NNArch}. For MTL, FNN and CNN classifiers feature 8,884  and 140,020 trainable parameters, respectively.

Multi-task classification accuracy results are shown in Table~\ref{tab:multitaskCNN} for CNN and Table~\ref{tab:multitaskFNN} for FNN. In these results, transmissions of both Device 1 and Device 2 as well as both legitimate and rogue devices are used. When we compare the same case in single task classification results reported in Table~\ref{tab:threstasksCNN} and Table~\ref{tab:threstasksFNN}, MTL improves classification accuracy in general, when CNN remains more successful than FNN in classification tasks.

\begin{table}[]
\footnotesize
    \centering
    \caption{Multi-task classification accuracy with CNN.}
    \label{tab:multitaskCNN}
    \begin{subtable}[t]{0.4\textwidth}
    \centering
    \caption{Task 1 of classifying `Device 1' vs. `Device 2'.}
    \begin{tabular}{l||l}
    \hline
    Pr\big($L_{\text{pred}}$ = $L_{\text{true}}$\big)  & 0.9476  \\ 
     Pr\big($L_{\text{pred}}$ = `Device 1' $\vert$ $L_{\text{true}}$ = `Device 1'\big) &  0.9325 \\ 
      Pr\big($L_{\text{pred}}$ = `Device 2' $\vert$ $L_{\text{true}}$ = `Device 2'\big) & 0.9627 \\ \hline
    \end{tabular}
    \vspace{0.25cm}
    \end{subtable}

    \begin{subtable}[t]{0.4\textwidth}
    \centering
    \caption{Task 2 of classifying `legitimate' vs. `rogue' devices.}
    \begin{tabular}{l||l}
    \hline
    Pr\big($L_{\text{pred}}$ = $L_{\text{true}}$\big)  & 0.9785 \\ 
     Pr\big($L_{\text{pred}}$ = `legitimate' $\vert$ $L_{\text{true}}$ = `legitimate'\big) & 0.9940 \\ 
     Pr\big($L_{\text{pred}}$ = `rogue' $\vert$ $L_{\text{true}}$ = `rogue'\big) & 0.9630  \\ \hline 
    \end{tabular}
    \vspace{0.5cm}
    \end{subtable} 

    \caption{Multi-task classification accuracy with FNN.}
    \label{tab:multitaskFNN}
    \begin{subtable}[t]{0.4\textwidth}
    \centering
    \caption{Task 1 of classifying `Device 1' vs. `Device 2'.}
    \begin{tabular}{l||l}

    \hline
    Pr\big($L_{\text{pred}}$ = $L_{\text{true}}$\big) & 0.9201 \\  
     Pr\big($L_{\text{pred}}$ = `Device 1' $\vert$ $L_{\text{true}}$ = `Device 1'\big) &  0.9138 \\
     Pr\big($L_{\text{pred}}$ = `Device 2' $\vert$ $L_{\text{true}}$ = `Device 2'\big) &  0.9264  \\ \hline
    \end{tabular}
    \vspace{0.25cm}
    \end{subtable}
    
    \begin{subtable}[t]{0.4\textwidth}
    \centering
    \caption{Task 2 of classifying `legitimate' vs. `rogue' devices.}
    \begin{tabular}{l||l}

    \hline
    Pr\big($L_{\text{pred}}$ = $L_{\text{true}}$\big) & 0.9763 \\ 
     Pr\big($L_{\text{pred}}$ = `legitimate' $\vert$ $L_{\text{true}}$ = `legitimate'\big) & 0.9792  \\ 
     Pr\big($L_{\text{pred}}$ = `rogue' $\vert$ $L_{\text{true}}$ = `rogue'\big) & 0.9724\\ \hline 
    \end{tabular}
    \end{subtable} 
    
\end{table}

\section{Adversarial Attacks} \label{sec:attack}
\subsection{Untargeted Attack on Single-Task Classification} \label{sec:untargeted}
FGSM method is chosen to generate adversarial examples since it is fast and computationally efficient \cite{goodfellow2015explaining}. It requires only a single gradient computation to generate adversarial examples. While there are other advanced, iterative defense methods available such as Projected Gradient Descent (PGD), Basic Iterative Method (BIM), Momentum Iterative Method (MIM), Carlini \& Wagner (C\&W) and DeepFool, since our goal is to demonstrate the feasibility of adversarial attacks on single-task and multi-task LoRa signal analysis, FGSM remains as an effective approach. 
This one-step attack strategically crafts adversarial perturbations by capitalizing on gradient information derived from the loss function concerning the input samples. For the untargeted attack, the objective of FGSM is to amplify the model's loss by altering the input data in the direction specified by the gradient sign. 

FGSM generates an adversarial perturbation for each input sample by computing the gradient of the loss function concerning the input data using model backpropagation and establishing the direction in which the input requires perturbation, by taking the sign of the gradient. A small magnitude ($\epsilon$) is used to proportionally scale the gradient sign to govern the intensity of the perturbation. The choice of $\epsilon$ serves as a parameter influencing the balance between the attack's strength and the detectability of the perturbation.

This attack adds the scaled perturbation to the original input sample, accomplished through element-wise addition. It is crucial to ensure that the resulting adversarial sample remains within the acceptable range of values, dictated by the data type and controlled by the adversary's maximum transmit power. In the untargeted adversarial attack, the goal is to determine a perturbation $\delta$ in the presence of a DNN model with parameters $\theta$, an input sample $x$, and its true label $y$. Formally, this optimization can be expressed as
\begin{equation}
\max_{\delta} L_i(x+\delta, y, \theta)  \label{eq:attack}
\end{equation}
for Classifier $i$ that is trained for Task $i=1,2$ subject to
\begin{enumerate}[label={C}{{\arabic*}}:]
\item $\|\delta\|_p \leq \epsilon_{max}$: the magnitude of perturbation $\delta$ is upper-bounded by $\epsilon_{max}$.
\item $x+\delta$ stays within the acceptable input range based on the device's transmit power and phase shift.
\end{enumerate}

Since solving the non-convex optimization problem in (\ref{eq:attack}) poses a significant challenge, FGSM addresses this by linearizing the loss function concerning the input perturbation since neural networks are hypothesized to be too linear to resist linear adversarial perturbations \cite{goodfellow2015explaining}. We can linearize the cost function around the current value of $\theta$, obtaining an optimal max-norm
constrained perturbation of $\delta$. 
An adversarial example $x_{\text{adv}}$ is crafted by manipulating the input in a way that amplifies the loss function relative to the true label. To achieve this for Classifier $i=1,2$, the loss function $L_i(x, y, \theta)$ is computed with respect to the true label $y$ for the input $x$. Subsequently, the gradient of the loss function is determined concerning the input, denoted as $\nabla_x L_i(x, y, \theta)$. The normalization of this gradient involves obtaining the sign of its elements. The sign of the gradient is scaled by a small perturbation magnitude $\epsilon$. The perturbation for the untargeted attack on Classifier $i=1,2$ is computed as
\begin{equation}
\delta = \epsilon \: \text{sign} \big(\nabla_x L(x, y, \theta)\big).
\label{eq:singlepertubation}
\end{equation} 
The adversarial example $x_{\text{adv}}$ is created by incorporating the perturbation into the original input, resulting in $x_{\text{adv}} = x + \delta$. 
An adversary transmits perturbation signal such that the LoRa receiver receives $x_{\text{adv}}$ and its DNN model is potentially deceived into producing incorrect classifications for any label. The attack success probability (ASP) refers to the probability of incorrectly classifying input signals without specifying a specific target label. We assess the ASP as a function of the perturbation-to-signal ratio (PSR), which dictates the upper limit on the magnitude of the perturbation denoted by $\epsilon$.

The adversarial perturbations generated for one model may exhibit a tendency to generalize and maintain effectiveness across multiple models \cite{papernot2016transferability}. This transferrability may occur because the vulnerabilities targeted by adversarial attacks are not exclusive to a specific model. Instead, they can be present in multiple models due to shared characteristics in their decision boundaries or loss landscapes. The impact of surrogate models and wireless channels on transferability has been explored in \cite{surrogate} within the context of a single task. In this paper, we assess how adversarial attacks transfer between  Classifiers 1 and 2 for Tasks 1 and 2. For that purpose, we consider individual and joint perturbation scenarios:
\begin{enumerate}
    \item `Perturbation for Classifier $i$' is the perturbation computed by (\ref{eq:singlepertubation}) to maximize the loss of Classifier $1$ for Task $i$. 
    \item `Perturbation for Classifier 1+2' is the perturbation computed according to 
\begin{equation} \label{eq:hybrid}
\delta = \epsilon \: \text{sign} \bigg( \sum_{i=1}^2 \gamma_i \nabla_x L_i(x, y_i, \theta_i) \bigg)
\end{equation} 
to maximize the weighted loss of Classifiers 1 and 2, where $L_i$ is the loss function for classifier $i$ with parameters $\theta_i$ and $\gamma_i$ is the weight for Classifier $i$, $i=1,2$ (such that $0 \leq \gamma_i \leq 1$ and $\gamma_1+\gamma_2= 1$) and. We set $\gamma_1=\gamma_2=0.5$ for numerical results. 
\end{enumerate}

\begin{figure}[t]
\vspace{-0.15cm}
	\centering
	\includegraphics[width=0.935\columnwidth]{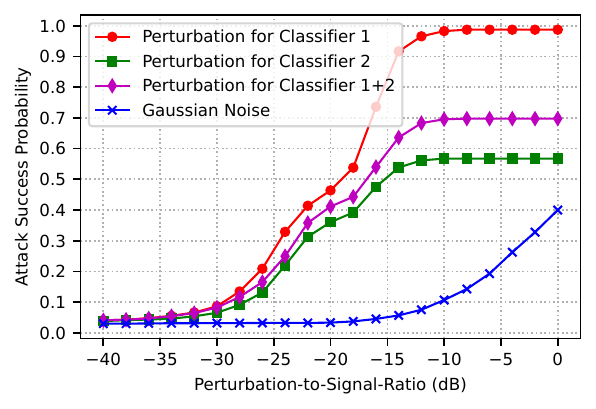}
	\caption{Untargeted attack performance on CNN classifier 1.}
	\label{fig:class1CNN}
        \vspace{0.25cm}
	\centering
	\includegraphics[width=0.935\columnwidth]{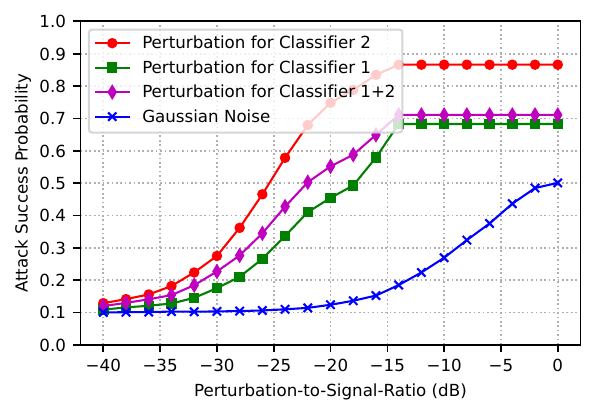}
	\caption{Untargeted attack performance on CNN classifier 2.}
	\label{fig:class2CNN}
\end{figure}

\begin{figure}[t]
\vspace{-0.15cm}
	\centering
	\includegraphics[width=0.935\columnwidth]{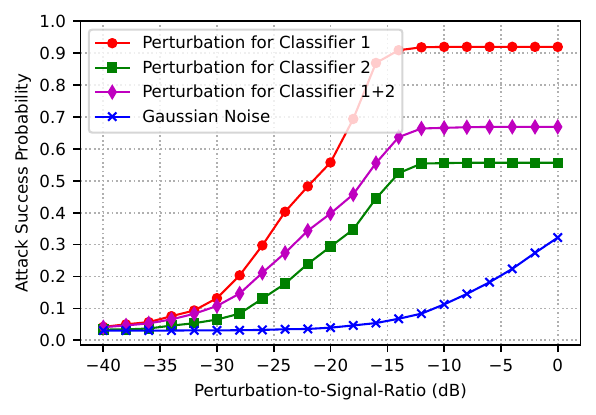}
	\caption{Untargeted attack performance on FNN classifier 1.}
	\label{fig:class1FNN}
        \vspace{0.25cm}
	\includegraphics[width=0.935\columnwidth]{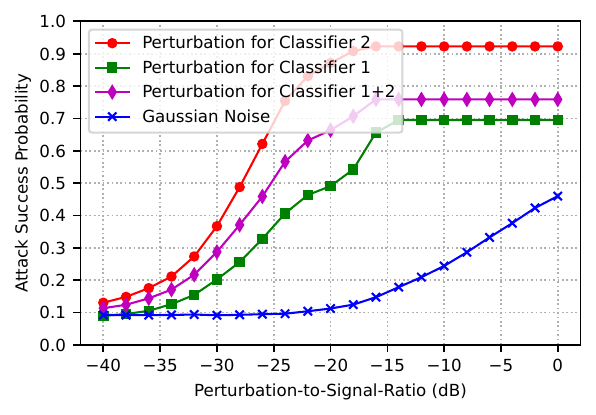}
	\caption{Untargeted attack performance on FNN classifier 2.}
	\label{fig:class2FNN}
\end{figure}

The untargeted ASPs on CNN-based Classifiers 1 and 2 are shown in Figs.~\ref{fig:class1CNN} and~\ref{fig:class2CNN}, respectively. Similarly, the untargeted ASP for the FNN-based Classifiers 1 and 2 are shown in Figs.~\ref{fig:class1FNN} and~\ref{fig:class2FNN}, respectively. As the benchmark, we assess the efficacy of adding Gaussian noise as a perturbation signal to the input samples. Overall, the goal is to evaluate the effects of small variations in test-time inputs, in terms of Gaussian noise and adversarially-crafted perturbations, and show when such variations become effective.
In all these results, the ASP is measured as a function of PSR in dB. Ideally, the PSR value should be small such as below $0$dB for the attack not to be detected easily. 
Our results show that `Perturbation for Classifier $i$' can effectively fool Classifier $i$ under an untargeted attack on Task $i=1,2$. 
On the other hand, the attack performance, namely the ASP, drops when `Perturbation for Classifier $i$' is used to fool Classifier $j$, where $j \neq i$, i.e., the classifier under attack is not the same as the classifier for which the perturbation is derived. The level of this drop in the ASP depends on the classifier under attack. Specifically, the ASP drops more for the attack on Classifier 1 compared to the attack on Classifier 2. 
By employing a hybrid attack on 'Classifier 1+2,' the impact of this mismatch is balanced concerning the transferability of untargeted attacks. This attack utilizes a shared perturbation that is broadcast in a single transmission, achieving a high ASP against each of the two classifiers. In every case, the use of Gaussian noise as a perturbation signal remains ineffective in causing a major drop in classifier accuracy. We also note higher ASP for the attacks on the CNN classifier compared to the attacks on the FNN classifier.

\subsection{Targeted Attack on Single-Task Classification}
\begin{figure}[ht]
\vspace{-0.15cm}
	\centering
	\includegraphics[width=0.935\columnwidth]{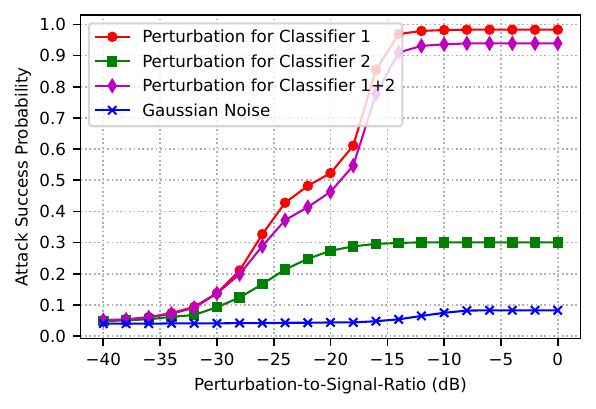}
	\caption{Targeted attack performance on CNN classifier 1.}
	\label{fig:targetCNN}
        \vspace{0.25cm}
	\includegraphics[width=0.935\columnwidth]{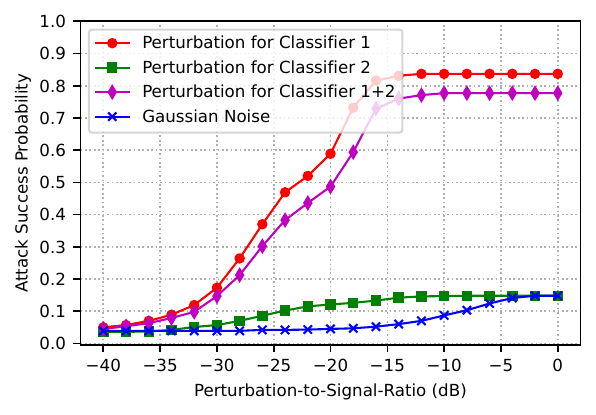}
	\caption{Targeted attack performance on FNN classifier 1.}
	\label{fig:targetFNN}
\end{figure}

\begin{figure}[ht]
\vspace{-0.15cm}
	\centering
	\includegraphics[width=0.935\columnwidth]{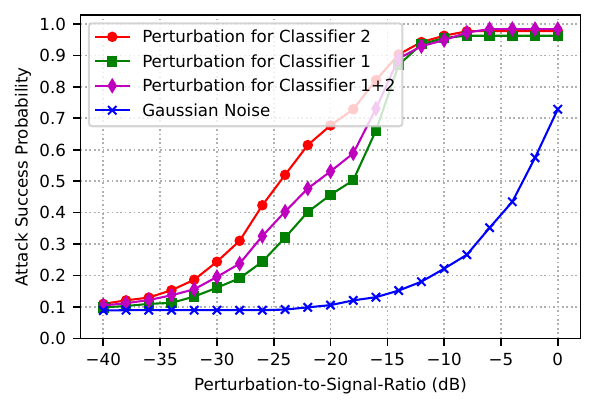}
	\caption{Targeted attack performance on CNN classifier 2.}
	\label{fig:targetCNN2}
        \vspace{0.25cm}
	\includegraphics[width=0.935\columnwidth]{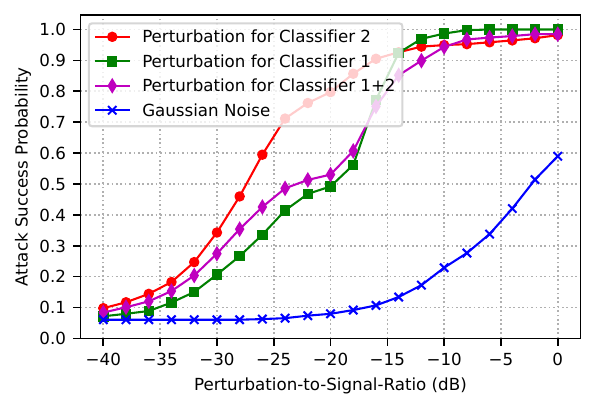}
	\caption{Targeted attack performance on FNN classifier 2.}
	\label{fig:targetFNN2}
\end{figure}

The optimization formulation for a targeted adversarial attack aims to determine a perturbation that maximizes the loss of the classifier under attack specifically for a desired target class. Given a DL model with parameters $\theta$, an input sample $x$, a desired target label $y_{\text{target}}$, and its true label $y$, the goal of the targeted attack based on FGSM is to find a perturbation $\delta$ that minimizes the loss function $L_i$ for Classifier $i=1,2$ such as
\begin{equation}
\min_{\delta} L(x+\delta, y_{\text{target}}, \theta) \label{eq:targeted_attack}
\end{equation}
subject to the same conditions C1 and C2 of the optimization problem in (\ref{eq:attack}). The optimization problem in (\ref{eq:targeted_attack}) is solved by FGSM that generates an adversarial example $x_{\text{adv}}$ by perturbing the input in the direction that minimizes the loss function with respect to the target label $y_{\text{target}}$. The adversarial example $x_{\text{adv}}$ is generated by adding the perturbation 
\begin{equation}
\delta = - \epsilon \: \text{sign}\big(\nabla_x L(x, y_{\text{target}}, \theta)\big)  
\end{equation}
to the original input such that $x_{\text{adv}} = x + \delta$. The resulting adversarial example $x_{\text{adv}}$ fools the model into incorrect misclassification for $y_{\text{target}}$. The ASP for targeted attack is the probability of classifying input signals from another label to $y_{\text{target}}$. We consider two targeted attacks: (i)  attack on Classifier 1 with $y_{\text{target}}$ = `Device 1' (fooling Classifier 1 classifying Device 2 signals to Device 1) (ii) attack on Classifier 2 with $y_{\text{target}}$ = `rogue' (fooling Classifier 2 classifying rogue signals as legitimate, namely  bypassing an RF fingerprinting-based authentication. Figs.~\ref{fig:targetCNN} and \ref{fig:targetFNN} show the ASP for the targeted attack on Classifier 1 based on CNN and FNN, respectively, when the perturbation is determined for Classifier 1 or 2, as well as by the hybrid attack (`Classifier 1+2') that uses the perturbation from (\ref{eq:hybrid}) with weights $\gamma_1=\gamma_2=0.5$. Similarly, Figs.~\ref{fig:targetCNN2} and \ref{fig:targetFNN2} show the ASP for the targeted attack on Classifier 2 based on CNN and FNN, respectively.

The targeted attack achieves high ASP when utilizing the perturbation that is specifically crafted for the intended classifier. However, when there is a discrepancy between the targeted model and the model for which the perturbation was derived, a decline in attack success is observed for both DNN types and the discrepancy is much more for the attack on Classifier 1. On the contrary, in the case of a hybrid attack (`classifier 1+2'), the impact of the mismatch on attack transferability is effectively mitigated, enabling the adversary to employ a common perturbation that can be disseminated to achieve a notably high ASP against each classifier (CNN or FNN). Again, the use of Gaussian noise as perturbation signal remains ineffective.

Regarding the time complexity, we measured the training and inference time for the single task classifiers. When model type is FNN, the average training time is 25.8225 sec and 25.8674 sec for Classifier 1 and 2, respectively. The average testing time is 0.1099 sec and 0.11 sec for Classifier 1 and 2, respectively. When model type is CNN, average training time is 38.1928 sec and 38.4808 sec for Classifier 1 and 2, respectively. The average testing time is 0.1312 sec and 0.1322 sec for Classifier 1 and 2, respectively.

\subsection{Untargeted Attack on Multi-Task Classification}
\begin{figure}[ht]
\vspace{-0.15cm}
	\centering
	\includegraphics[width=0.935\columnwidth]{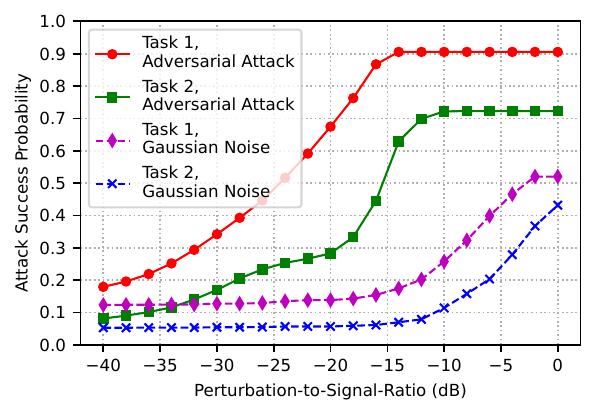}
	\caption{Untargeted attack performance on CNN multi-task classifier.}
	\label{fig:multitask_untargeted_CNN}
        \vspace{0.25cm}
	\includegraphics[width=0.935\columnwidth]{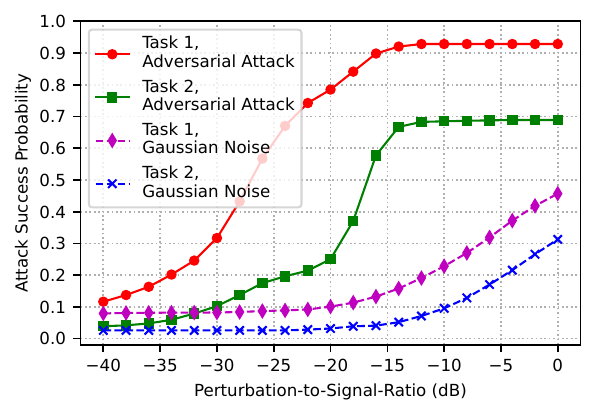}
	\caption{Untargeted attack performance on FNN multi-task classifier.}
	\label{fig:multitask_untargeted_FNN}
  \vspace{-0.3cm}
\end{figure}

By applying the FGSM method to the multi-task classifier, the adversarial perturbation for the untargeted attack is computed as
\begin{equation}
\delta = \epsilon \: \text{sign}\bigg(  \sum_{i=1}^2 \gamma_i \nabla_x L(x, y_i, \theta) \bigg),
\label{eq:jointpertubation}
\end{equation} 
where $\gamma_i$ is the weight for the contribution from Task $i$'s loss function to the perturbation (we assume $\gamma_i=0.5$, $i=1,2$, for numerical results. The adversarial example $x_{\text{adv}}$ is generated by adding the perturbation to the original input such that $x_{\text{adv}} = x + \delta$).  Figs.~\ref{fig:multitask_untargeted_CNN} and \ref{fig:multitask_untargeted_FNN} show the ASP for the untargeted attacks against CNN and FNN multi-task classifiers, respectively. 
 Overall, adversarial attacks are highly effective against multi-task classifiers (similar to the hybrid attack against single-task classifiers) and provide considerably better results compared to the attacks with Gaussian noise as the perturbation with increasing PSR. 

\subsection{Targeted Attack on Multi-Task Classification}
The goal is to flip labels of Tasks 1 and 2 to target labels $y_{\text{target},1}$ and $y_{\text{target},2}$.  The adversarial example $x_{\text{adv}}$ is generated by adding the perturbation 
\begin{equation}
\delta = - \epsilon \: \text{sign}\bigg(  \sum_{i=1}^2 \gamma_i \nabla_x L(x, y_{\text{target},i}, \theta) \bigg)    
\end{equation}
to the original input such that $x_{\text{adv}} = x + \delta$. 

\begin{figure}[ht]
\vspace{-0.15cm}
	\centering
	\includegraphics[width=0.935\columnwidth]{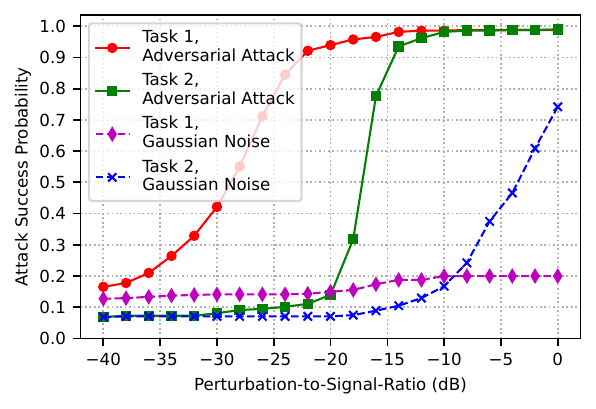}
	\caption{Targeted attack performance on CNN multi-task classifier.}
	\label{fig:multitask_targeted_CNN}        
        \vspace{0.25cm}
	\includegraphics[width=0.935\columnwidth]
        {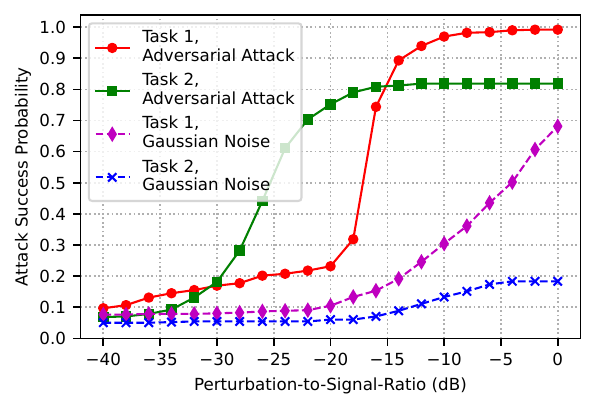}
	\caption{Targeted attack performance on FNN multi-task classifier.}
	\label{fig:multitask_targeted_FNN}
\end{figure}

For numerical results, we assume that the goal is to flip decisions from detecting rogue Device 2 to legitimate Device 1, i.e., $y_{\text{target},1}$ = `Device 1' and $y_{\text{target},2}$ = `legitimate'. Figs.~\ref{fig:multitask_targeted_CNN} and  \ref{fig:multitask_targeted_FNN} show the ASP for targeted attacks on CNN and FNN classifiers, respectively. The ASP increases quickly when the PSR increases. Overall, the ASP is higher when the DNN under attack is CNN compared to FNN, and reaches above $90\%$ for both Task 1 and Task 2, while the impact of using Gaussian noise as adversarial perturbation on classifier accuracy remains small. 

We measured the training and inference time for the multi-task classifiers. When model type is FNN, average training time is 26.2958 sec. The average testing time is 0.1182 sec. When model type is CNN, average training time is 40.7357 sec. The average testing time is 0.1354 sec.

\section{Defense against Adversarial Attacks} \label{sec:defense}
We use adversarial training as part of the DNN training process to make the DNN models robust against adversarial manipulations. Adversarial training aims to determine the model parameters $\theta$ for the DNN that can handle worst-case perturbations and achieve robustness against adversarial attacks \cite{madry2017towards}. Samples with adversarial perturbations are incorporated in the training set, expanding the dataset and exposing the model to adversarial instances. 

Given a training set of labeled examples $(x_i, y_i)$, where $x_i$ represents the input sample and $y_i$ represents the corresponding true label, adversarial training as a defense against untargeted attacks aims to determine model parameters $\theta$ that minimize the expected loss on adversarial examples as follows:
\begin{equation}
\min_{\theta} \mathbb{E}_{(x_i, y_i)} \left[\max_{\delta \in \mathcal{S}} L(x_i + \delta, y_i, \theta)\right],
\end{equation}
where $\delta$ represents the adversarial perturbation and $\mathcal{S}$ is the set of allowable perturbations subject to the conditions C1 and C2 of the optimization problem in (\ref{eq:attack}). The inner maximization term seeks the worst-case adversarial perturbation within the allowable set $\mathcal{S}$ that maximizes the loss on the perturbed example that is generated with FGSM. 

Table~\ref{tab:defense} shows the defense performance against the untargeted attack on classifier 1 when perturbation of -3dB PSR is determined for classifier 1, 2 and 1+2 (as discussed in Section~\ref{sec:untargeted}). The performance measures are the ASP before and after defense as well as the defense cost that is reflected by the accuracy after defense when there is no attack. For both CNN and FNN models, the defense is effective especially when the perturbation is determined either fully or partially for the target classifier 1 such that the attack success drops with defense while the effect on the classifier accuracy remains high with this defense even when there is no attack (i.e., the defense does not adversely affect the accuracy). 

Table~\ref{tab:multidefense} shows the defense performance against untargeted attack on multi-task classifier when the PSR is -3dB. In this case, the ASP considerably decreases when the defense is applied for both CNN and FNN classifiers.    

\begin{table}[ht]
\footnotesize
    \centering
    \caption{Defense performance for single-task classifiers.}
    \label{tab:device12_1}
    \begin{subtable}[t]{0.5\textwidth}
    \centering
    \caption{DNN type: CNN.}
    \label{tab:defenseCNN}
    \begin{tabular}{l||c|c|c}
    Perturbation for Classifier & 1 & 2 & 1+2 \\ 
    \hline
    ASP Before Defense &0.9870 & 0.5615 &  0.6940      \\   
    ASP After Defense & 0.0029 & 0.0275 & 0.0679  \\ 
    Accuracy without Attack & 0.9602 & 0.9634 & 0.9019 \\ \hline 
    \end{tabular}
    \vspace{0.6cm}
    \end{subtable}
    \begin{subtable}[t]{0.5\textwidth}
    \centering
    \caption{DNN type: FNN.}
    \label{tab:defenseFNN}
    \begin{tabular}{l||c|c|c}
   Perturbation for Classifier & 1 & 2 & 1+2 \\ 
    \hline
    ASP Before Defense & 0.9200 & 0.5545 & 0.6685  \\ 
    ASP After Defense & 0.0002 & 0.0554 & 0.0356  \\ 
    Accuracy without Attack & 0.9706 & 0.9435 & 0.9578 \\ \hline 
    \end{tabular}
    \end{subtable} \label{tab:defense}
    \end{table}

\begin{table}[ht]
\footnotesize
    \centering
    \caption{Defense performance for multi-task classifier.}
    \label{tab:device12_2}
    \begin{subtable}[t]{0.5\textwidth}
    \centering
    \caption{DNN type: CNN.}
    \label{tab:multidefenseCNN}
    \begin{tabular}{l||c|c}
    Attack on & Task 1 & Task 2 \\
    \hline
    ASP Before Defense & 0.9055 & 0.7225      \\   
    ASP After Defense & 0.0612 & 0.0601  \\ 
    Accuracy without Attack & 0.8942 & 0.9500  \\ \hline 
    \end{tabular}
    \vspace{0.5cm}
    \end{subtable}
    \begin{subtable}[t]{0.5\textwidth}
    \centering
    \caption{DNN type: FNN.}
    \label{tab:multidefenseFNN}
    \begin{tabular}{l||c|c}
    Attack on & Task 1 & Task 2 \\
    \hline
    ASP Before Defense & 0.9289 & 0.6890    \\   
    ASP After Defense & 0.1005 & 0.1236  \\ 
    Accuracy without Attack & 0.8921 &  0.9565\\ \hline 
    \end{tabular}
    \end{subtable} \label{tab:multidefense}
    \end{table}

\section{Conclusion} \label{sec:conclusion}
In this paper, we have addressed security concerns related to LoRa networks, which offer long-range and low-power communication for IoT applications. Our approach involves developing a DL framework for two critical tasks: distinguishing between LoRa devices and distinguishing between legitimate and spoofed LoRa signals (where KDE is adopted to generate spoofed signals by rogue devices). I/Q data from LoRa signals captured by SDRs are used to train the DNNs (CNN or FNN) for these tasks by considering either two separate classifiers or a single multi-task classifier. Then we investigated the impact of FGSM-based untargeted and targeted adversarial attacks on these LoRa signal classification tasks. Our findings demonstrated the high effectiveness of these attacks in significantly reducing the classification accuracy by using a common perturbation against two tasks captured by either two single-task classifiers, one for each of the two tasks, or a multi-task classifier. These results underscore the vulnerability of LoRa-based IoT systems, and stress the imperative need for robust defenses against such attacks. Finally,  we applied adversarial training to make the classifiers robust against the adversarial attacks, and showed the feasibility of the defense to ensure the reliable and secure operation of LoRa networks. The developed defense mechanism is effective against the adversarial examples generated by the FGSM method. It is not guaranteed to work against unseen attack types. Applying other defense methods such as defensive distillation  \cite{eed59049cc7f4604a5d58c4a28bc19f9}, randomized smoothing and certified defense \cite{cohen2019certified} and analyzing their effectiveness for unseen attack types can be studied as part of the future work.
\bibliographystyle{IEEEtran}
\bibliography{references}

\end{document}